# On Quantum Collapse as a Basis for the Second Law of Thermodynamics

Ruth E. Kastner [1]

[1] University of Maryland, College Park; rkastner@umd.edu

March 7, 2017

**Abstract:** It was first suggested by David Z. Albert that the existence of a real, physical non-unitary process (i.e., "collapse") at the quantum level would yield a complete explanation for the Second Law of Thermodynamics (i.e., the increase in entropy over time). The contribution of such a process would be to provide a physical basis for the ontological indeterminacy needed to derive the irreversible Second Law against a backdrop of otherwise reversible, deterministic physical laws. An alternative understanding of the source of this possible quantum "collapse" or non-unitarity is presented herein, in terms of the Transactional Interpretation (TI). The present model provides a specific physical justification for Boltzmann's often-criticized assumption of molecular randomness (*Stosszahlansatz*), thereby changing its status from an *ad hoc* postulate to a theoretically grounded result, without requiring any change to the basic quantum theory. In addition, it is argued that TI provides an elegant way of reconciling, via indeterministic collapse, the time-reversible Liouville evolution with the time-irreversible evolution inherent in so-called "master equations" that specify the changes in occupation of the various possible states in terms of the transition rates between them. The present model is contrasted with the Ghirardi-Rimini-Weber (GRW) 'spontaneous collapse' theory previously suggested for this purpose by Albert.

## 1. Introduction

Irreversible processes are described by the Second Law of Thermodynamics, the statement that entropy $S$ can never decrease for closed systems: $\frac{dS}{dt} \geq 0$. This law is corroborated ubiquitously at the usual macroscopic level of experience. However, there remains significant debate regarding exactly how it is that these commonplace irreversible processes arise from an ostensibly time-reversible level of description. Specifically, it is commonly assumed that the quantum level obeys only the unitary dynamics of the time-dependent Schrödinger equation, which is time-reversible. In addition, classical mechanics can be obtained as the small-wavelength limit of the unitary quantum evolution, as Feynman showed in his sum-over-paths approach [1].

Boltzmann originally introduced irreversibility into his "H-Theorem," a derivation of the Second Law, through his *Stosszahlansatz* (assumption of molecular chaos)[2]. This assumption consists of treating molecular and atomic state occupations as stochastic and independent before interactions but not necessarily after (for an in-depth account of this and related issues, Brown, Myrvold and Uffink [3]).

After objections from Loschmidt [4] and Zermelo [5], Boltzmann modified his understanding of the H-Theorem, concluding that he had demonstrated that the accessible thermodynamic macrostates corresponding to higher entropy were far more probable than those corresponding to lower entropy. As a number of researchers have noted (e.g., Lebowitz [6], Albert [7]) one can obtain macroscopic irreversibility from this consideration along with (i) a "statistical postulate" (defining a suitable probability measure over the portion of phase space corresponding to each macrostate of the system), and (ii) a postulate of a low-entropy past initial condition for the universe (termed the "Past Hypothesis" by Albert [7]). However, Penrose[8] expresses some concerns about the extreme fine-tuning required to satisfy condition (ii). Meanwhile, although Lanford [9] was widely credited with having derived irreversibility without relying on the *Stosszahlansatz*, Uffink and Valente [10] have cogently argued that in fact his theorem leads to no genuine irreversibility. Popescu et al [11] offer a unitary-only, quantum-level account of the Second Law, but it obtains as an approximation, and is arguably subject to the same circularity concerns regarding the emergent macroscopic 'pointer' basis that plague the unitary-only 'decoherence' program [12].

Thus, there remains some lack of consensus over the significance of the H-Theorem and the fundamental explanation of macroscopic irreversibility. One might argue that the arrow of time missing from the microscopic reversible laws is inserted through the two auxiliary postulates, especially (ii); we choose a low-entropy past rather than a low-entry future. Of course, this seems perfectly reasonable given our empirical experience, but it would appear to introduce an element of circularity into any putative explanation for our perceived temporal arrow. Regarding the statistical assumption (i), Sklar has noted that "[t]he status and explanation of the initial probability assumption remains the central puzzle of non-equilibrium statistical mechanics"[13].

Now, if it were an incontrovertible fact that all we have are reversible, deterministic laws at the microlevel, the account of the Second Law through postulates (i) and (ii) would presumably be adequate. Yet there is a possibility, as Albert first noted, that there is indeed a form of non-unitary evolution at the micro-level. If so, it becomes much more straightforward to demonstrate the occurrence of macroscopic irreversibility. This possibility is explored herein. The present treatment differs from that of Albert, who was working with the GRW 'spontaneous collapse' theory [14]. The latter includes an explicitly *ad hoc* change to the Schrödinger evolution in order to force collapse into the position basis. In contrast, the present proposal does not require an *ad hoc* change to the basic Schrödinger evolution in order to account for the stochastic behavior corresponding to 'coarse graining.' Rather, by taking into account the response of absorbers in a direct-action model of fields, it provides the physical referent for an expression already appearing in the mathematical formalism of quantum theory: the Von Neumann non-unitary measurement transition from a pure state to an epistemic mixed state. This feature will be discussed in Section 4.

It should perhaps be noted at the outset that the present author is aware that there is a very long history of discussion and debate, and a vast literature, on this topic. The current work makes no pretense of providing a comprehensive account of the history of the debate (a careful and thorough treatment can be found in the work of Uffink, Brown, and Myrvold cited above). Rather, it focuses on the narrow issue of proposing a specific alternative model as the source of the time-asymmetric 'coarse graining' required for a physically grounded account of entropy increase.

Let us now briefly review the basic problem.

**2. Reversible vs non-reversible processes**

Classical laws of motion are in-principle reversible with respect to time. There is a one-to-one relationship between an input *I* and an output *O*, where *I* and *O* are separated by a time interval *Δt*. If *Δt* is taken as positive, then *I* is the cause and *O* is the effect. If we reverse the sign of *Δt*, then the roles of the output and input are simply exchanged; the process can just as easily run backwards as forwards. The same applies to quantum processes described by the Schrödinger equation: the input and output states are linked in a one-to-one relationship by deterministic, unitary evolution.

Moreover, it is well established that the 'statistical operator' (density operator), *ρ,* applying to a quantum system obeys a unitary, time reversible dynamics, analogous to the Liouville equation for the phase space distribution of microstates in classical statistical mechanics. The general definition of the density operator (applying either to a pure or mixed state) is:

$$\rho = \sum_i P_i |\Psi_i\rangle\langle\Psi_i| \tag{1}$$

where $P_i$ is the probability that the system is in the pure state $|\Psi_i\rangle$, and the $P_i$ sum to unity. The states $|\Psi_i\rangle$ need not be orthogonal, so in general $\{|\Psi_i\rangle\}$ is not a basis.

From the Schrödinger equation and its adjoint, ones finds the time-evolution of $\rho$:

$$\frac{\partial \rho}{\partial t} = \frac{-i}{h}[H, \rho]$$

(2)

where $H$ is the Hamiltonian. It is important however to note that $\rho$ is not an observable; this is reflected in the sign difference between its time dependence and that of an observable $O$, which obeys $\frac{\partial O}{\partial t} = \frac{i}{h}[H, O]$. The significance of this point is that it is in-principle impossible to "observe" what state the system is in through any measurement. Of course, one can *prepare* any arbitrary state one wishes, but (as is well known, for example, in the spreading of wave packets) for any finite time after preparation, in general the status of the system relative to any property corresponding to a value of an observable is fundamentally uncertain. This provides a clue as to where the 'probability assumption' must enter for any quantum process that is well-defined relative to a particular observable (such as momentum)—and this point will be explored in what follows.

In contrast to the unitary evolution (2), non-unitary evolution such as that described by von Neumann's "Process 1," or measurement transition, is indeterministic [15]. An input pure state $I$ is transformed to one of many possible output states $O_i$, elements of a particular basis, with no causal mechanism describing the occurrence of the observed output state $O_k$.[1] The different possible outcomes are statistically weighted by probabilities $P_i$ according to the Born Rule. As a result of the measurement transition, the system is represented by a mixed state $\tilde{\rho}$. This one-to-many transition is inherently irreversible; once a final state occurs, the original state is not accessible to it through simple time reversal.

Of course, the status of the non-unitary measurement transition has long been very unclear. It is often thought of as epistemic in nature—i.e., describing only a Bayesian updating of an observer's knowledge. Such an epistemic view of quantum measurement has its own interpretive challenges, which we will not enter into here;[2] but it also can provide no ontological basis for the observed asymmetry described by the Second Law. On the other hand, if the measurement transition is a real (indeterministic) physical process, it is clearly a candidate for the ontological introduction of stochastic randomness--describable by probabilities such as those in master equations--and the resulting irreversibility described by the Second Law. In fact, Von Neumann himself showed that his 'Process 1' is irreversible and always entropy-increasing [15]. However, he seemed to have veered away from using that fact in deriving the Second Law, because he thought of the measurement transition as dependent on an external perceiving consciousness, and as such not a real physical process.

**3. Standard Approaches to the Second Law; "Smuggling In" Non-unitarity**

A typical 'derivation' of the Second Law begins with unitary evolution to obtain the basic transition rates between various states, but ends up with a master equation from which one finds that the time rate of change in entropy is always positive (or zero for equilibrium). We'll consider this seeming paradox in what follows. First, recall that a master equation relates the change in the probability $P_i$ that a system is in state $|i>$ to the transition rates $R_{ij}$ between that state and other states $|j>$. Specifically:

$$\frac{dP_i}{dt} = \sum_j R_{ij} P_j - R_{ji} P_i \equiv [M] P_i$$

(3)

---

[1] Of course, hidden variables theories attempt to provide a causal mechanism by 'completing' quantum theory, but here we consider quantum mechanics as already complete and simply in need of a direct-action(transactional) interpretation.

[2] For example, the Pusey-Rudolph-Barrett theorem [16] rules out most statistical interpretations of the quantum state.

where [M] is the 'master operator.' Each diagonal element of [M] is the negative of the sum of all the off-diagonal elements in the same column (which are all positive). This property gives rise to a decaying exponential time-dependence, yielding an irreversible tendency to an equilibrium state, independently of the initial state of the system. As an illustration, consider a simple example in which the transition probabilities $R_{ij}$ between states 1 and 2 are both ½. The solutions for $P_i$ (i=1,2) will be:

$$P_1(t) = \frac{1}{2} + \frac{P_1(0) - P_2(0)}{2} e^{-2t}$$

$$P_2(t) = \frac{1}{2} + \frac{P_2(0) - P_1(0)}{2} e^{-2t}$$

(4)

We can see from the above that with increasing time, the second term, containing the initial state information, approaches zero and one is left with the equilibrium distribution $P_1(t_\infty) = P_2(t_\infty) = \frac{1}{2}$. Thus, the equilibrium distribution is the final result, without regard to the initial state. Determinism is broken.

Let us now examine how irreversibility 'sneaks in' despite the time-reversible evolution represented by the Liouville equation. Irreversibility appears through the use of master equations, such as (3), employing transition rates between the occupied states $|i>$. First, recall that the Von Neumann entropy $S_{VN}$ is defined in terms of the density operator in a basis-independent way as:

$$S_{VN} = -Tr(\rho ln\rho) \qquad (5)$$

However, in order to employ the back-and-forth 'detailed balance' between states needed for master equations, one must work within a particular basis $\{|i>\}$ corresponding to transitions between the relevant states. So rather than work with the density operator, one typically uses a diagonal density matrix:

$$\tilde{\rho} = \sum_i P_i |i\rangle\langle i| \qquad (6)$$

where $P_i$ is the probability that the system is in state $|i\rangle$. In that basis, (5) becomes

$$S = -\sum_i(P_i ln P_i) \qquad (7)$$

This form, the Shannon entropy, is proportional (by a factor of Boltzmann's constant $k_B$) to the Gibbs entropy, which is still conserved in any unitary, deterministic process. Therefore, entropy cannot increase unless there is an element of randomness along with the underlying Liouville (deterministic) evolution. The latter corresponds to the 'coarse graining' or 'blurring' of the fine-level trajectories resulting from Liouville evolution. The question now obviously arises: how is this 'blurring' related to the non-unitarity inherent from master equations such as (3)? In the context of classical systems, the traditional answer is that in order to obtain the relevant rates used in master equations, one has to deal in practice with an approximate description, owing to the enormous complexity of the macroscopic system under study. This is thought of as 'throwing out information'—an epistemic interpretation of the 'coarse-graining'. And at the classical level, that is the only possible source of the 'blurring'.

However, at the quantum level, in order to define the rates of change $dP_i/dt$ used in master equations, any phase coherence in quantum states is lost. This is a loss of information that can be understood in ontological (rather than epistemic) terms, in contrast to the classical level, if there is ongoing real non-unitary projection into the basis *{i}*; i.e., repeated transformations of any initial pure state to an epistemic (proper) mixed state. Thus, if such non-unitary projection actually occurs during the evolution of a given system (such as a gas), then its entropy does increase, despite the governing deterministic Hamiltonian dynamics; the phase-space conserving evolution of the Liouville equation is physically broken at the micro-level.

Such a model is proposed in what follows. Again, it should be noted that the present account of the source of the coarse-graining is distinct from that proposed by Albert [7], in that it specifically underlies the already-codified von Neumann measurement transition, and does not involve any change to the Schrödinger evolution. Rather than changing the basic theory, the present model provides a specific physical account of the transition from the standard Schrödinger unitary evolution to the von Neumann epistemic mixed state. This non-unitary process occurs whenever the system transitions from one of its microstates to another, provided such transitions arise from inelastic processes (such as thermal interactions). The present proposal also differs from the GRW approach in that it treats the conserved quantities (e.g. energy and momentum) as privileged observables, rather than treating position $X$ as privileged, which is not tenable at the relativistic level (since position is not a relativistically well-defined observable). The latter issue is discussed in Section 4.

## 3. The Transactional Interpretation

Before turning to the specifics of TI, it is worth noting that Einstein himself posited a fundamental quantum irreversibility associated with the particle-like aspect of light. Since it is the latter that accounts for the measurement transition and accompanying irreversibility in the TI model, let us revisit his comments on this point:

> In the kinetic theory of molecules, for every process in which only a few elementary particles participate (e.g., molecular collisions), the inverse process also exists. But that is not the case for the elementary processes of radiation. According to our prevailing theory, an oscillating ion generates a spherical wave that propagates outwards. The inverse process does not exist as an elementary process. A converging spherical wave is mathematically possible, to be sure; but to approach its realization requires a vast number of emitting entities. *The elementary process of emission is not invertible*. In this, I believe, our oscillation theory does not hit the mark. Newton's emission theory of light seems to contain more truth with respect to this point than the oscillation theory since, first of all, the energy given to a light particle is not scattered over infinite space, but remains available for an elementary process of absorption. [17]; emphasis added]

The above comments were made when it was well-established that light has both a wave and particle aspect. In order to explain well-known interference effects of light, a wave model of photon emission is needed. However, as Einstein himself pointed out via the empirical phenomenon of the photoelectric effect, light is absorbed in particle-like, discrete quanta. He is thus noting that, for a single quantum, all the energy represented by an isotropically propagating emitted wave (i.e., a superposition of all wave vectors **k** of a given energy) ends up being delivered to only a single absorbing system; thus the process acquires a final anisotropy (i.e., a specific wave vector **k**) not present initially. The latter is a feature of the particle-like aspect of light, and that is what makes the process non-invertible. (This microscopic origin of irreversibility was also pointed out by Doyle [18].) As we will see, TI acknowledges both a wavelike and particlelike aspect to light, and it is the latter that brings about the irreversibility, just as Einstein noted.

*3.1. Background*

The Transactional Interpretation was first proposed by Cramer [19] based on the Wheeler-Feynman direct-action theory of classical fields [20,21]. Its recent development by the present author [23-27] is based on the fully relativistic direct-action quantum theory of Davies [28,29]. In view of this relativistic development, the model is now referred to as the Relativistic Transactional Interpretation (RTI). It should perhaps be noted at the outset that TI is not considered a 'mainstream' interpretation, since its underlying model of fields—the direct-action theory—has historically been viewed with various degrees of skepticism. Nevertheless, despite the counterintuitive nature of the model, which includes advanced solution to the field equations, there is nothing technically wrong with it. (See [27] for why Feynman's abandonment of his theory was unnecessary.) Moreover, no less a luminary than John A. Wheeler was recently attempting to resurrect the direct-action theory in the service of progress toward a theory of quantum gravity. It's worth quoting from that paper here, in order to allay any concerns about the basic soundness of the model:

> [WF] swept the electromagnetic field from between the charged particles and replaced it with "half-retarded, half advanced direct interaction" between particle and particle. It was the high point of this work to show that the standard and well-tested force of reaction of radiation on an accelerated charge is accounted for as the sum of the direct actions on that charge by all the charges of any distant complete absorber. Such a formulation enforces global physical laws, and results in a quantitatively correct description of radiative phenomena, without assigning stress-energy to the electromagnetic field. ([30], p. 427)

Thus, there is no technical reason to eliminate the direct-action approach, and every reason to reconsider it in connection with such longstanding problems as the basis of the Second Law.

*3.2 Measurement in the Transactional Interpretation*

An overview of the Transactional Interpretation (henceforth "TI") is provided in [26]. To briefly review: according to the absorber theory, the basic field propagation is time-symmetric, containing equal parts retarded and advanced fields. When such a field is emitted, absorbers are stimulated to respond with their own time-symmetric field, which is exactly out of phase with the emitted field. This gives rise to a real retarded field directed from the emitter to the absorber, and this is what accounts for the loss of energy by a radiating charge.[3]

TI is a "collapse" interpretation. That is, in general, many absorbers M will respond to an emitted field in this way, but since the field is quantized, in the case of a field corresponding to N photons, only N of the M absorbers can actually receive the conserved quantities (energy, momentum, etc.) contained in the emitted field. The choosing of one or more "winning" absorbers for receipt of the photon(s), as opposed to all the other possible sites for energy transfer (i.e., the many absorbers "losing the competition") is what corresponds to "collapse." It is a completely indeterministic matter as to which absorber(s) will actually receive the real energy. The response of absorbers, leading to collapse, is what breaks the linearity of the Schrödinger evolution and allows TI to physically define the measurement transition without reference to an outside "observer."

Let us now focus on the quantitative aspects of the TI account of the measurement transition. For present purposes it is sufficient to recall that according to TI, the usual quantum state or 'ket' $|\Psi\rangle$ is referred to as an 'offer wave' (OW), or sometimes simply 'offer' for short. The unfamiliar and counter-intuitive aspect of the direct action theory is inclusion of the solution to the complex conjugate (advanced) Schrodinger equation; this is the dual or 'brac,' $\langle X_i|$, describing the response of one or more absorbers $X_i$ to the component of the offer received by them. The advanced responses of absorbers are termed 'confirmation waves' (CW).[4] Specifically, an absorber $X_k$ will receive an offer wave component $\langle X_k|\Psi\rangle|X_k\rangle$ and will respond with a matching adjoint confirmation $\langle \Psi|X_k\rangle\langle X_k|$. The product of the offer/confirmation exchange is a weighted projection operator, $\langle X_k|\Psi\rangle \langle\Psi|X_k\rangle |X_k\rangle\langle X_k| = |\langle X_k|\Psi\rangle|^2 |X_k\rangle\langle X_k|$. Clearly, the weight is the Born Rule, and this is how TI provides a physical origin for this formerly *ad hoc* rule. When one takes into account the responses of all the other absorbers { $X_i$ }, what we have is the von Neumann measurement transition from a pure state to a mixed state $\tilde{\rho}$:

$$|\Psi\rangle \rightarrow \tilde{\rho} = \sum_i |\langle\Psi|X_i\rangle|^2 |X_i\rangle\langle X_i| \qquad (8)$$

---

[3] The emitted field is the time-symmetric solution to the inhomogenous wave equation, and therefore has a discontinuity at the source. In contrast, the field resulting from the combination of the retarded component from the emitter and the (inverse phase) advanced absorber response is a retarded source-free 'free field,' i.e., a solution to the homogenous wave equation.

[4] As their names indicate, both of these objects are wavelike entities—specifically, they are deBroglie waves.

In the absence of absorber response, the emitted offer wave (OW), $|\Psi\rangle$, is described by the unitary evolution of the time-dependent Schrodinger equation. Equivalently, in terms of a density operator $\rho = |\Psi\rangle\langle\Psi|$, its evolution can be described by its commutation with the Hamiltonian, as in (2).[5] However, once the OW $|\Psi\rangle$ prompts response(s) $\langle X_i|$ from one or more absorbers {$X_i$}, the linearity of this deterministic propagation is broken, and we get the non-unitary transformation (8).

Thus, according to TI, absorber response is what triggers the measurement transition. (Precise quantitative, though indeterministic, conditions for this response are discussed in [24].) It is the response of absorbers that transforms a pure state density operator $\rho$ to a mixed state $\tilde{\rho}$, diagonal with respect to the basis defined by the absorber response, as shown in (8). And in fact it is here that the "probability assumption" enters in a physically justified manner, since the system is now physically described by a set of random variables (the possible outcomes) subject to a Kolmogorov (classical) probability space. All phase coherence is lost.

The second step in the measurement transition is collapse to one of the outcomes $|X_k\rangle\langle X_k|$ from the set of possible outcomes {i} represented by the weighted projection operators $|\langle\Psi|X_i\rangle|^2 |X_i\rangle\langle X_i|$ in the density matrix $\tilde{\rho}$ above. This can be understood as a generalized form of spontaneous symmetry breaking, a weighted symmetry breaking: i.e., actualization of one of a set of possible states where in general the latter may not be equally probable. This is where Einstein's particle-like aspect enters. For example, an emitted isotropic (spherical) electromagnetic offer wave is ultimately absorbed by only one of the many possible absorbers that responded to it with CWs. The transferred quantum of electromagnetic energy acquires an anisotropy: a single directional momentum corresponding to the orientation of the 'winning' absorber (which is called the *receiving absorber* in RTI). All the other possible momentum directions are not realized. The anisotropic directedness of the actualized wave vector **k** corresponds to the particle-like aspect or photon.[6] Since the latter process exchanges a determinate quantity of energy/momentum—a photon Fock state -- the energy/momentum basis can be understood as distinguished. We return to the latter issue when we consider the relativistic level, in Section 4 below.

In view of the above, it is apparent that a physically real measurement transition naturally leads to probabilistic behavior accompanied by loss of phase coherence, thereby instantiating the 'coarse-graining' required for entropy increase. If interacting systems are engaging in continual emission/absorption events constituting 'Process 1,' these non-unitary processes quickly destroy any quantum coherence that might arise. Between confirming interactions (these being inelastic as opposed to elastic), component systems may be described by deterministic (unitary) evolution; but with every inelastic interaction, that evolution is randomized through the underlying quantum non-unitarity. Moreover, any receiving absorber becomes correlated with the emitter through the delivery of the emitted photon--which acquires a specific wave vector **k** corresponding to that absorber (i.e., as noted above, the spherically emitted offer wave collapses to only one momentum component). The emitter loses a quantum of energy/momentum and the absorber gains the same, leaving an imprint of the interaction (at least in the short term), which thus establishes the time-asymmetric conditions of the *Stosszahlansatz*. Thus, the time-asymmetric statistical description that Boltzmann assumed in order to derive the Second Law is justified, based on a real physical process.

---

[5] However, TI is best understood in the Heisenberg picture, in which the observables carry the explicit time dependence. Also, at the relativistic level, OW and CW are generated together in a mutual process; we never really have an OW without one or more CW (this is the quantum relativistic equivalent of the 'light-tight box' condition of the absorber theory). The unitary evolution characterizes force effects on the OW and CW in between their emitting/absorbing systems.

[6] Of course, in this respect, 'particle-like' does not mean having a localized corpuscular quality. Rather, the 'particle' is a discrete quantum of energy/momentum. The directionality of the final received photon momentum is what localizes the expanding spherical wave to a particular final individual absorbing gas molecule, resulting in approximate localization of the transferred photon.

Once again, the above proposal differs from that of Albert, in that it does not change the basic theory; rather, it simply provides missing physical referents for computational processes that are already part of the theory, yet which are usually not interpreted physically (the Born Rule and the Von Neumann measurement transition). According to the present proposal, the crucial missing ingredient in the standard account is a physical model of the non-unitary measurement transition, which accompanies all inelastic microstate changes. This is what yields a physical explanation for why thermal interactions are correctly characterized by randomness and time-asymmetric correlations, just as Boltzmann assumed.

## 4. The Relativistic Level: Further Roots of the Arrow of Time

At the deeper, relativistic level of TI as it has been developed ([22-27]), the generation of absorber response (i.e. a confirmation) is itself a stochastic process described (in part) by coupling amplitudes between fields. For example, the random Poissonian probabilistic description of the decay of an atomic electron's excited state is understood in the RTI picture as reflecting a real ontological indeterminacy in the generation of both an offer and confirmation for the photon emitted. Details of the transactional model of the inherently probabilistic nature of atomic decays and excitations are given in [24]. The same basic picture applies to other kinds of decays (i.e. of nuclei or composite quanta), since all such decays occur due to coupling among the relevant fields.

Considering the relativistic level also allows us to identify a basic source of temporal asymmetry corresponding to that pointed out by Einstein above. In the direct-action theory, the state of the quantum electromagnetic field resulting from absorber response to the basic time-symmetric propagation from an emitter is a Fock state [24]. These correspond to 'real photons'; they are quantized, positive-energy excitations of the field. Such states can be represented by the action of creation operators $\hat{a}^\dagger$ on the vacuum state of the field. E.g., a single photon state of momentum $k$ is given by:

$$|k\rangle = \hat{a}^\dagger{}_k |0\rangle$$

Meanwhile, the confirming response, a 'brac' or dual ket $\langle k|$, can be represented as the annihilation operator $\hat{a}_k$ acting to the left on the dual vacuum, i.e.:

$$\langle k| = \langle 0|\hat{a}_k$$

The relevant point is that there is an intrinsic temporal asymmetry here: *a field excitation must be created before it can be destroyed* (or, equivalently, responded to by an absorbing system). This seemingly obvious and mundane fact is actually a crucial ingredient in the origin of the temporal arrow: any emission must precede the corresponding absorption. An emission event therefore must always be in the past relative to its matching absorption event. This is simply because one cannot destroy something that does not exist: a thing must first exist in order to be destroyed. The basic relativistic field actions of creation and annihilation therefore presuppose temporal asymmetry. This asymmetry is reflected in the distinctly different actions of the creation and annihilation operators on the vacuum state:

$$\hat{a}^\dagger{}_k |0\rangle = |k\rangle; \quad \text{whereas} \quad a_k |0\rangle = 0.$$

Thus, if one tries to annihilate something that doesn't exist, one gets no state at all—not even the vacuum state.

The above is why the indeterministic collapse to one out of many possible outcomes (typically, one of many possible wave vectors for a photon transferred from one gas molecule to another) also yields a temporal directionality—i.e., an arrow of time. The chosen outcome always corresponds to the transfer of a quantum of energy (and momentum, angular momentum, etc). Energy is the generator of temporal displacement, and since a quantum must be created before it is destroyed, the energy transfer always defines a temporal orientation *from* the emitter (locus of creation) *to* the absorber (locus of annihilation). Moreover, the delivered energy is always positive, corresponding to a positive temporal increment.[7]

Thus, the present model contains within it a natural source for the temporal arrow without having to appeal to large-scale cosmological conditions as an additional postulate, resolving Penrose's concern [8]. For any local inelastic interaction to occur, all one needs is an excited atom or molecule (potential emitter) and one or more atoms/molecules capable of receiving the associated quantum of energy (potential absorbers). Any resulting transaction, conveying a photon from the emitter to one of the responding absorbers, carries with it an arrow of time at the micro-level, even if the macroscopic state of the overall system does not change. In particular, this means that a sample of gas in thermal equilibrium, apparently manifesting no temporally-oriented behavior, still contains micro-level temporally oriented processes.

In addition, it should be noted that taking the relativistic level into account provides a physically grounded way of 'breaking the symmetry' of the various possible observables. It is commonly supposed that there is no fundamental way of identifying any distinguished observable (or set of observables), but that view arises from taking the nonrelativistic theory as a complete and sufficient representation of all the relevant aspects of Nature, when it is not: Nature is relativistic, and the nonrelativistic theory is only an approximate limit. At the relativistic level of quantum field theory, there is no well-defined position observable, since position state vectors are non-orthogonal; this fact provides a natural reason to consider the spacetime parameters as ineligible for a privileged basis.

In any case, at all levels, there is a fundamental distinction between the spacetime description and the energy/momentum description: the spacetime indices parametrize a symmetry manifold, while energy and momentum are conserved physical quantities (they are the Noether currents generating the symmetry properties of the spacetime manifold). In that sense, the two sorts of descriptors (spacetime vs energy/momentum) are very different physically. Moreover, there is no time observable, even at the nonrelativistic level. Thus, even apart from relativistic considerations, there are sound physical reasons to demote the spacetime quantities to mere parameters and to treat the energy/momentum basis as privileged. This approach is in contrast to unitary-only treatments, which typically help themselves to features of our macroscopic experience (e.g., apparent determinacy of position or at least quasi-localization of systems of interest) in order to specify preferred observables and/or Hilbert Space decompositions, rather than providing a specific theoretical justification for these choices at the fundamental microscopic level. In the approach presented here, quasi-localization arises because of the collapse to a particular spatial momentum singling out the receiving absorber, even though the distinguished basis, describing the transferred physical quantity, is energy/momentum. In the view of the present author, the current model of non-unitarity is thus an improvement over the GRW model, which treats spacetime parameters as privileged.

## 5. Conclusion

It has been argued that if the non-unitary measurement transition of Von Neumann is a physically real component of quantum theory, then the representation of the system(s) under study by proper mixed states, subject to a probabilistic master equation description relative to a distinguished basis, becomes physically justified. This rectifies a weakness in the usual approach, which helps itself to the convenient basis and accompanying probabilistic description (effectively

---

[7] The direct action theory is subject to a choice of boundary conditions for the superposition of the time-symmetric fields from emitters and absorbers leading to the free field component needed for real (on-shell) energy propagation. The choice discussed herein corresponds to Feynman propagation. It is possible to choose Dyson rather than Feynman propagation, but the resulting world is indistinguishable from our own; the definition of 'negative' vs 'positive' energy is just a convention in that context. For further discussion of this issue, in terms of Gamow vectors and resulting microscopic proper time asymmetry, see [31].

Pauli's "random phase assumption")[32] as a 'for all practical purposes' approximation. However, the utility of a probabilistic expression for calculational purposes does not constitute theoretical justification for the probabilistic description, which is needed in order for the 'coarse-graining' and resulting entropy increase to describe what is physically occurring in a system. Once we have that justification, through real non-unitary collapse, we have the microscopic irreversibility needed to place the H-theorem on sound physical footing.

According to the TI account of measurement, a quantum system undergoes a real, physical non-unitary state transition based on absorber response, which projects it into a Boolean probability space defined with respect to the observable being measured (typically energy in the context of thermodynamics). Thus the system's probabilistic description by random variables is justified; the non-unitary measurement transition can be understood as the physical origin of the 'initial probability assumption' referred to as puzzling by Sklar. In this model, it ceases to be an assumption and can be seen as describing a physical feature of Nature. In effect, Boltzmann was completely correct about the *Stosszahlansatz*, even though he could not explain why in classical terms.

In addition, the relativistic level of TI (referred to as RTI) provides a physical reason for the directionality of the irreversibility inherent in the measurement transition at the micro-level, thereby establishing an arrow of time underlying the Second Law. In this respect, the microscopic arrow of time becomes a component of the explanation for the increase in the entropy of closed systems towards what we call "the future," without the need for an additional postulate of a cosmological low-entropy past.

The present model has been contrasted with the GRW model proposed by Albert, as follows: it does not require any change to the basic Schrödinger evolution, but simply provides a physical account of the non-unitary measurement transition previously formalized by Von Neumann. In addition, the present model takes conserved physical quantities (energy, momentum, etc.) as the preferred eigenbasis of collapse, rather than position as in the GRW theory. This takes into account the fact that position is not an observable at the relativistic level (and time is not an observable at any level). It is also in accordance with the naturally occurring microstates of thermodynamical systems, in which the molecular components are described by distributions over their energies and momenta (i.e., the Boltzmann distribution applies to energies, not positions).

Acknowledgments. The author is indebted to Prof. David Albert and two anonymous referees for valuable comments.

**References**

1. Feynman, R. P.; Hibbs, A. R. *Quantum Mechanics and Path Integrals*. New York: McGraw-Hill, 1965.
2. Boltzmann,L. "Weitere Studien über das Wärmegleichgewicht unter Gasmolekülen." *Sitzungsberichte Akademie der Wissenschaften* **1872**, 66, 275-370.
3. Brown, H.R., Myrvold, W., and Uffink. J. Boltzmann's H-theorem, its discontents, and the birth of statistical mechanics. Studies in History and Philosophy of Modern Physics 40 (2009) 174–19
4. Loschmidt, J. *Sitzungsber. Kais. Akad. Wiss. Wien, Math. Naturwiss. Classe* **1876**, 73, 128–142.
5. Zermelo, E. Uber enien Satz der Dynamik und die mechanische Warmetheorie. *Annalen der Physik,* **1896,** 57, 485-94.
6. Lebowitz, J. L Time's Arrow and Boltzmann's Entropy. *Scholarpedia* **2008**. 3(4), 3448.
7. Albert, D. Z. *Time and Chance*. Cambridge: Harvard University Press, 2000, pp. 150-162.
8. Penrose, R. *The Emperor's New Mind*. Penguin Books, 1989, pp 339-345.
9. Lanford, O. E. On the derivation of the Boltzmann equation. *Asterisque* **1976**, 40, 117–137.
10. Uffink, J. and Valente, G. Lanford's Theorem and the Emergence of Irreversibility, *Fnd. Phys.* **2015**, 45: 404-438.
11. Popescu, S., Short, A., and Winter, A. Entanglement and the foundations of statistical mechanics. *Nature Physics* 2, 754-758 (2006).

12. Kastner, R. E. Einselection of Pointer Observables: The New H-Theorem? *Studies in History and Philosophy of Modern Physics 48*: 56–58. (2014). Preprint version: http://philsci-archive.pitt.edu/10757/
13. Sklar, L. "Philosophy of Statistical Mechanics", The Stanford Encyclopedia of Philosophy (Fall 2015 Edition), Edward N. Zalta (ed.), URL <http://plato.stanford.edu/archives/fall2015/entries/statphys-statmech/>. Accessed on 12 December 2016.
14. Ghirardi, G.C., Rimini, A., and Weber, T. "A Model for a Unified Quantum Description of Macroscopic and Microscopic Systems". *Quantum Probability and Applications, L. Accardi et al. (eds), Springer, Berlin*, **1985**.
15. Von Neumann, J. *Mathematical Foundations of Quantum Mechanics*. (Trans: Beyer, R.T.) Princeton: Princeton University Press, 1955, pp. 347-445.
16. Pusey, M., Barrett, J., and Rudolph,T. On the Reality of the Quantum State. *Nature Physics* 8, 475–478 (2012).
17. Einstein, A. "On the Development of Our Views Concerning the Nature and Constitution of Radiation." *Einstein Collected Papers* **1909**, 2, p.387.
18. Doyle, R. O. "The continuous spectrum of the hydrogen quasi-molecule." *J. Quant. Spectrosc. Radiat. Transfer* **8**, 1555–1569 (1968).
19. Cramer J. G.   ``The Transactional Interpretation of Quantum Mechanics." *Reviews of Modern Physics* 58, 647-688, 1986.
20. Wheeler, J.A. and R. P. Feynman, "Interaction with the Absorber as the Mechanism of Radiation," Reviews of Modern Physics, 17, 157–161 (1945).
21. Wheeler, J.A. and R. P. Feynman, "Classical Electrodynamics in Terms of Direct Interparticle Action," Reviews of Modern Physics, 21, 425–433 (1949).
22. Kastner, R. E. "The New Possibilist Transactional Interpretation and Relativity." Fnd. of Phys. **2012**, 42, 1094-1113.
23. Kastner, R. E. *The Transactional Interpretation of Quantum Mechanics: The Reality of Possibility*. Cambridge: Cambridge University Press, 2012.
24. Kastner, R. E. "On Real and Virtual Photons in the Davies Theory of Time-Symmetric Quantum Electrodynamics," *Electronic Journal of Theoretical Physics* **2014**, 11(30), 75–86.
25. Kastner, R. E. "The Emergence of Spacetime: Transactions and Causal Sets," in Licata, I. (Ed.), *Beyond Peaceful Coexistence*. Singapore: World Scientific, 2016.
26. Kastner, R. E. "The Transactional Interpretation: an Overview," *Philosophy Compass* **2016,** 11(12), 923-932.
27. Kastner, R. E. "Antimatter in the direct-action theory of fields," **2016**, *Quanta 5(1)*, pp. 12-18. arXiv:1509.06040
28. Davies, P. C. W. Extension of Wheeler-Feynman Quantum Theory to the Relativistic Domain I. Scattering Processes,"   *J. Phys. A: Gen. Phys*. **1971**, 6, p. 836
29. Davies, P. C. W."Extension of Wheeler-Feynman Quantum Theory to the Relativistic Domain II. Emission Processes," J. Phys. A: Gen. Phys. **1972**, 5, p. 1025.
30. Wesley, D. and Wheeler, J. A., "Towards an action-at-a-distance concept of spacetime," In *Revisiting the Foundations of Relativistic Physics*: *Festschrift in Honor of John Stachel,* Boston Studies in the Philosophy and History of Science (Book 234), A. Ashtekar et al, eds.; Kluwer Academic Publishers, **1972**, pp. 421-436.
31. Gaioli, F.H., Garcia-Alvarez, E.T. & Castagnino, M.A. "The Gamow Vectors and the Schwinger Effect*."* Int J Theor Phys **1997**, 36, p. 2371. doi:10.1007/BF02768930
32. Pauli, W. *Festschrift zum 60sten Geburtstag A. Sommerfelds*. Hirzel, Leipzig **1928**, p. 30.